\documentclass[journal]{IEEEtran}
\usepackage{ulem}
\usepackage{amsmath,amssymb}

\usepackage{setspace}
\usepackage{graphicx}
\usepackage{tikz}
\usepackage{color}
\usepackage{lipsum,adjustbox}
\usetikzlibrary{decorations}
\usepackage{pgfplots,siunitx}
\usetikzlibrary{decorations.pathreplacing}
\usetikzlibrary{shapes,arrows,shadows}
\usetikzlibrary{calc,positioning}
\usepackage{tikz,tkz-euclide}

\newcommand{\argmax}{\mathop{\mathrm{argmax}}}

\pgfplotsset{compat=1.3}

\usetikzlibrary{calc}
\usepackage{subfig}
\usetikzlibrary{external}
\tikzsetfigurename{tikz_compiled} 

\begin{document}
\title{A Machine Learning-Based Detection Technique for Optical Fiber Nonlinearity Mitigation}
\author{Abdelkerim~Amari, Xiang~Lin, Octavia~A.~Dobre, Ramachandran~Venkatesan, Alex~Alvarado 
\thanks{Abdelkerim~Amari was with the Faculty of Engineering and Applied Science, Memorial University,  St. John's, Canada. He is currently with the Information and Communication Theory Lab, Signal Processing Systems Group, Department of Electrical Engineering, Eindhoven University of Technology, The Netherlands.
Email: a.amari@tue.nl.}
\thanks{Xiang~Lin, Octavia~A.~Dobre, Ramachandran~Venkatesan, are with the Faculty of Engineering and Applied Science, Memorial
University,  St. John's, Canada.}
\thanks{Alex~Alvarado is with the Information and Communication Theory Lab, Signal Processing Systems Group, Department of Electrical Engineering, Eindhoven University of Technology, The Netherlands.}
\thanks{This work was supported by Atlantic Canada Opportunities Agency (ACOA) and Research Development Corporation (RDC) Canada, and by the Netherlands Organisation for Scientific Research (NWO) via the VIDI Grant ICONIC (project number 15685).}
}

\maketitle
\begin{abstract}
 We investigate the performance of a machine learning classification technique, called the Parzen window, to mitigate the fiber nonlinearity in the context of dispersion managed and dispersion unmanaged systems. The technique is applied for detection at the receiver side, and deals with the non-Gaussian nonlinear effects by designing improved decision boundaries. We also propose a two-stage mitigation technique using digital back propagation and Parzen window for dispersion unmanaged systems. In this case, digital back propagation compensates for the deterministic nonlinearity and the Parzen window deals with the stochastic nonlinear signal-noise interactions, which are not taken into account by digital back propagation. A performance improvement up to $0.4$ dB in terms of Q factor is observed.
\end{abstract}

\IEEEpeerreviewmaketitle


\section{Introduction}
\IEEEPARstart{F}{iber} nonlinearity mitigation has been considered as a key technology to increase the optical system capacity. Several digital signal processing (DSP) techniques have been proposed to compensate for the nonlinear distortions in the optical link, as reviewed in \cite{winzer}--\cite{am1}. 
Machine learning techniques have recently received significant attention as promising approaches to deal with such effects. These techniques have been applied as detectors at the receiver side \cite{ml4}--\cite{ml7}, and also as channel model-based compensation algorithms \cite{ml2}--\cite{ml3}.

Machine learning-based detectors provide two main advantages. Firstly, they can  partially mitigate both deterministic fiber nonlinearities and stochastic nonlinear signal-amplified spontaneous emission (ASE) noise interactions. Secondly, they do not require the knowledge of the optical link parameters, which makes them well-suited for dynamic optical networks.

Multiple machine learning-based detectors have been proposed in the context of dispersion unmanaged (DUM) and dispersion managed (DM) systems, such as support vector machines (SVM) \cite{ml4}, K-means clustering \cite{ml5}, and the K-nearest neighbors algorithm \cite{ml6}. The main idea of machine learning-based detectors is to design improved nonlinear decision boundaries more adapted to the nonlinear fiber channel. Thus, the nonlinear distortions such as nonlinear non-Gaussian noise can be mitigated.

In this letter, we propose a machine learning-based classification technique, known as the Parzen window (PW) classifier \cite{pw,pwknn}, to mitigate the non-Gaussian nonlinear effects. The PW classifier is applied as a detector at the receiver side.
 We show that a performance improvement in terms of the $Q$ factor is observed when applying the PW classifier to both DUM and DM systems. 
 
 \begin{figure}[!tbph]
	\centering		
		\includegraphics[width=0.95\linewidth]{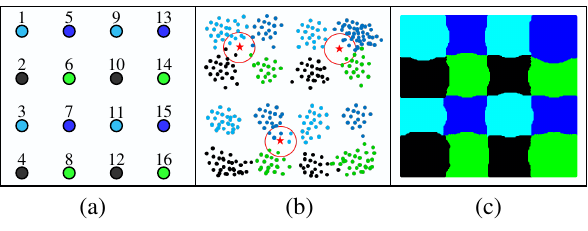}
	\caption{ PW principle: (a) Tx training symbols; (b) Rx detection based on training symbols; (c) Rx decision regions.} 
\label{fig:PW}
\end{figure}

 \begin{figure*}[!tbph]
 \vspace{-0.2cm}
	\centering		
		\includegraphics[width=0.99\linewidth]{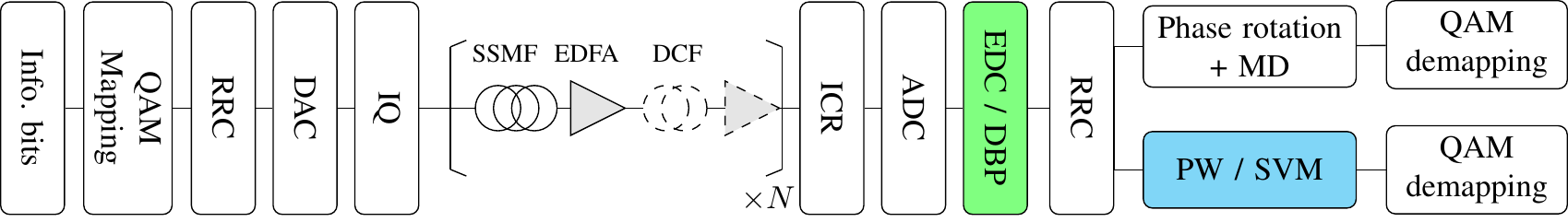}
		\caption{Transmission diagram: IQ: In-phase and quadrature modulator, ICR: Integrated coherent receiver. Dashed lines represent the partly-considered components.}
\label{fig:SIM}
\end{figure*}

Furthermore, we complement the PW classifier with digital back propagation (DBP) \cite{ef1}, and propose a two-stage fiber nonlinearity mitigation. DBP is used to compensate for the deterministic nonlinear effects. Then, the PW-based detector is applied to deal with the nonlinear non-Gaussian signal-ASE noise interactions. In DUM systems, the two-stage nonlinearity mitigation using  DBP and the PW classifier increases the performance in comparison with DBP, used with the classic minimum distance (MD)-based detector.

\section{Parzen Window Principle }\label{sec:PW}

Machine learning techniques are widely investigated in the context of optical communication systems, as discussed in \cite{ml1}--\cite{ml1--}. In particular, machine learning-based classification techniques have been proposed as detectors to deal with the nonlinear effects. In this context, we propose the PW, which is a machine learning nonparametric classification technique based on supervised learning. It is inherently a multi-class technique and can be applied for multi-level modulation formats without adaptation. This is unlike other proposed machine learning techniques, such as SVM \cite{ml6}, which is a binary classifier; multiple SVMs are required for the detection of high-order modulations. In addition, PW does not require an offline training process, like SVM and artificial neural networks.

The main idea of PW is to associate a label to each symbol, and then classify it at the receiver based on labeled training data. The principle of the PW-based detection is depicted in Fig.~\ref{fig:PW}, where $16$-QAM is used as example, and  the closest neighboring constellation points are represented using different colors. Fig.~\ref{fig:PW}~(a) shows the transmitter side where the $M$-QAM constellation points $\{s_1,s_2,\ldots,s_M\}$ are labeled to $M$ clusters. We denote the label of each cluster $s_m$ by $m \in\{1,2,\ldots,M\}$. At the beginning of the transmission, $T$ training symbols $x_k$ (and their corresponding labels $m_k$) with $k=1,2,\ldots,T$ are generated. This training data is followed by $N-T$ testing data symbols $x_k, k=T+1,T+2,\ldots,N$. The $N$ symbols $[x_1,x_2,\ldots,x_N]$ are transmitted over the optical channel. At the receiver side, for each received testing symbol $y_k$, one Euclidean distance between $y_k$ and each received training symbol is calculated. Thus, for each $k=T+1,\ldots,N$, $T$ Euclidean distances denoted by $D_{k,t}=|y_k-y_{t}|$, where $t=1,\ldots,T$ are calculated.

The decision rule of the PW technique depends on two parameters: a window size $R$ and a window function $f_{k,t}$. Both parameters should be optimized and adapted to the classification problem. Since the data is distributed in a 2-dimensional plane, a circle with radius $R$ centered around the testing symbol is employed as the window shape. This process is schematically shown in Fig.~\ref{fig:PW}~(b), where three testing symbols (stars) are shown together with the labeled training symbols (colored dots). Furthermore, we use a kernelized window function, in which the closest training points to the testing data have the highest significance, namely,
\begin{align}\label{f}
f_{k,t} &= 
\begin{cases} 
\frac{1}{D_{k,t}} & \mbox{if } D_{k,t}\leq R \\ 
0 & \mbox{otherwise}
\end{cases},
\end{align}
where $t=1,\ldots,T$. 

The last step in the classification process is to compute a metric $L_{k,m}$ for each possible transmitted symbol (cluster) $m$. This metric is calculated by adding up all the contributions of $f_{k,t}$ in \eqref{f} for each training cluster, i.e.,
\begin{align}\label{:}
L_{k,m}=\sum_{\substack{t=1\\x_t=s_m}}^{T} f_{k,t}, \qquad m=1,\ldots,M.
\end{align}
The estimated cluster is then the one with the largest metric, i.e., $\hat{m}_k=\argmax_{m\in\{1,\ldots,M\}} L_{k,m}$, and thus, the estimated symbol is $\hat{x}_k=s_{\hat{m}_k}$, with $k=T+1,\ldots,N$. An example of decision regions generated by the PW-based detection is depicted in Fig.~\ref{fig:PW}~(c).

The use of the inverse of the Euclidean distance as a weight for the window function $f$ improves the performance of the PW technique. It also avoids the particular case of having two clusters with the exact same metric $L_m$. A similar idea was considered in \cite{ml6}, by using square Euclidean distances.


\section{Simulation Setup and Results}\label{sec:simus}

The performance of the PW classifier is investigated by numerical simulation. In this \mbox{simulation}, we consider a single-channel dual-polarization configuration. We focus on the intra-channel fiber nonlinear effect and neglect the stochastic laser phase noise and polarization mode dispersion.
 The simulation setup is shown in Fig.~\ref{fig:SIM}. We compare the performance of the PW-based detector in combating the fiber nonlinearity with MD and SVM-based detections.

We consider $16$-QAM and $64$-QAM DM and DUM systems, in which the total bit rate is $224$ Gbps.  For the DUM system, the transmission link consists of multi-span standard single mode fiber (SSMF) with an attenuation coefficient $\alpha=0.2~\mathrm{dB\cdot km^{-1}}$, a dispersion parameter $D=16 ~\mathrm{ps \cdot nm^{-1} \cdot km^{-1}}$, and a nonlinear coefficient \mbox{$\gamma=1.4~ \mathrm{W^{-1} \cdot km^{-1}}$}. 
An erbium-doped fiber amplifier (EDFA) with a $5.5$ dB noise figure and $16$ dB gain is used at each span of $80$ km. When a DM system is considered, an additional EDFA and a dispersion-compensated fiber, with full chromatic dispersion (CD) compensation, are deployed at each span.
A root-raised cosine (RRC) filter with a roll-off factor $\rho = 0.1$ is employed for spectrum shaping and the analog-to-digital converter (ADC) works at twice the symbol rate. $1000$ symbols are used as training symbols for $16$-QAM modulation and $2000$ symbols in case of $64$-QAM modulation. $2^{14}$ symbols are used as testing data.

The DSP at the receiver consists of CD compensation, and deterministic fiber nonlinearity mitigation via DBP (if used). After that, an RRC matched filter is applied. Finally, PW-based detection is performed before QAM demapping and error counting. When applying the PW classifier, the phase rotation compensation is not required because the signal detection is based on the labeled training symbols. However, the phase compensation is carried out by using training sequence for the minimum distance-based detection. 

\begin{figure}[!tbph]
	\centering		
		\includegraphics[width=0.93\linewidth]{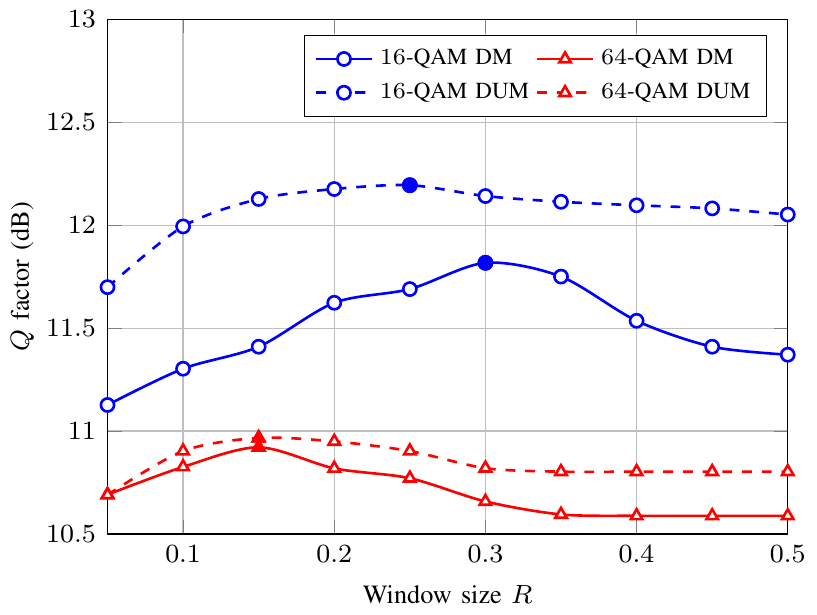}
		\caption{$Q$ factor vs. window size $R$ for DM and DUM systems.}	
		\label{fig:A}
\end{figure}

The performance of the PW depends on the window size $R$, which should be optimized based on the transmission parameters, such as the input power and the transmission distance. For example, in Fig.~\ref{fig:A}, we plot the $Q$ factor, calculated as in \cite{am1}, versus the PW size $R$ for  $16$-QAM at $800$ km transmission distance, and $64$-QAM at $240$ km, at optimal input power and for both DM and DUM systems. The optimal window sizes are $R = 0.3$ and $R = 0.15$ for $16$-QAM and $64$-QAM DM system, and $R = 0.25$ and $R = 0.15$ for $16$-QAM and $64$-QAM DUM system, respectively.

\begin{figure}[!tbph]
	\centering		
		\includegraphics[width=0.93\linewidth]{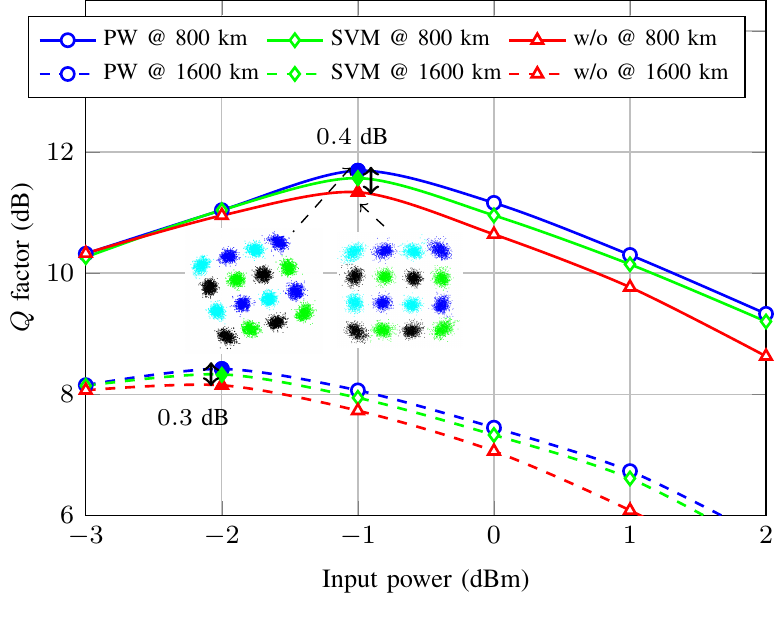}
		\caption{$Q$ factor vs. input power for DM $16$-QAM.}	
		\label{fig:B}
\end{figure}
We firstly focus on the performance evaluation of the PW-based detector in the DM system. Fig.~\ref{fig:B} shows the $Q$ factor versus the input power for $16$-QAM at $800$ km  and $1600$ km. We compare the performance of the PW with MD preceded by phase compensation, referred to as (w/o) in the figure, and SVM. At optimal input powers, the PW technique improves the performance by about $0.4$ dB  and $0.3$ dB in comparison with MD for $800$ km and $1600$ km, respectively. A Q factor increase of about $0.1$ dB is observed when compared to SVM. $1000$ symbols are used in the training process of the SVM to determine the model parameters, which is the same as the number of training symbols for PW. Increasing the number of training symbols for both PW and SVM can increase the performance, but results in higher complexity of the algorithms.

In Fig.~\ref{fig:B}, we also show the constellation diagrams of the detected symbols at optimum input power $-1$ dBm for PW and MD at $800$ km. These constellation plots emphasize that machine learning techniques, and in particular PW, can detect the signal without the need of phase rotation compensation. This is due to the design of new decision boundaries depending only on the training symbols.

\begin{figure}[!tbph]
	\centering		
		\includegraphics[width=0.93\linewidth]{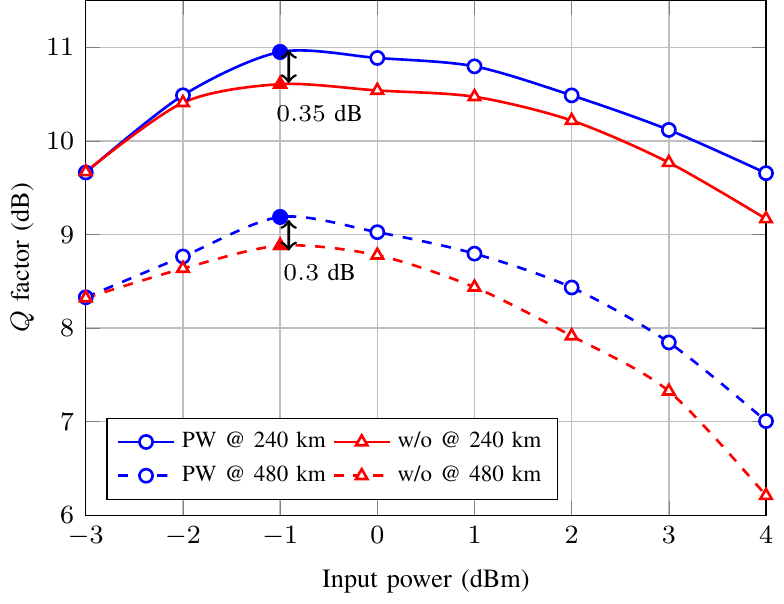}
		\caption{$Q$ factor vs. input power for DM $64$-QAM.}	
		\label{fig:C}
\end{figure}

In Fig.~\ref{fig:C}, we plot the $Q$ factor performances for $64$-QAM at $240$ km and $480$ km. At optimal input power, the PW-based detector increases the performance in comparison with the MD-based detector, by about $0.35$ dB and $0.3$ dB for $240$ km and $480$ km, respectively. In the linear regime, PW-based detector provides similar performance to the MD-based detector, while a significant improvement is observed in the nonlinear regime, due to the increased nonlinear non-Gaussian noise.  

\begin{figure}[!tbph]
	\centering		
		\includegraphics[width=0.93\linewidth]{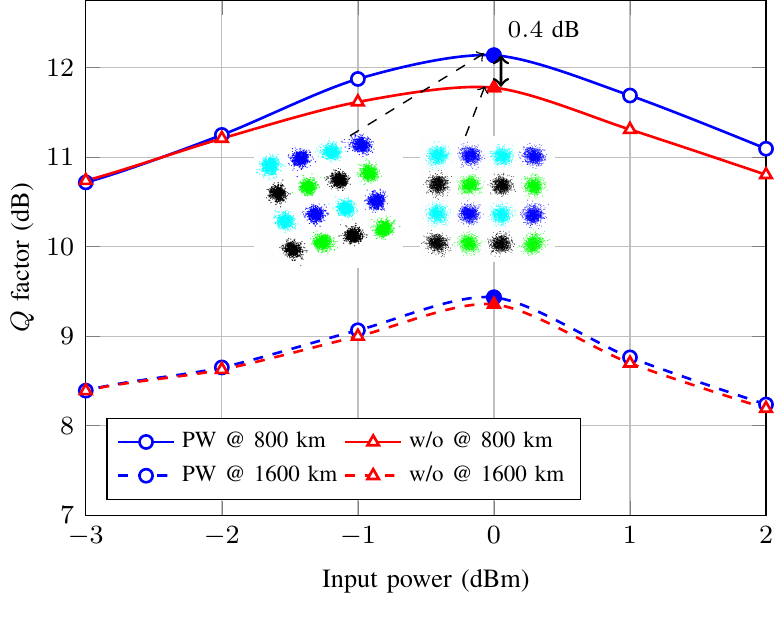}
		\caption{$Q$ factor vs. input power for DUM $16$-QAM.}	
		\label{fig:BB}
\end{figure}

\begin{figure}[!tbph]
	\centering		
		\includegraphics[width=0.93\linewidth]{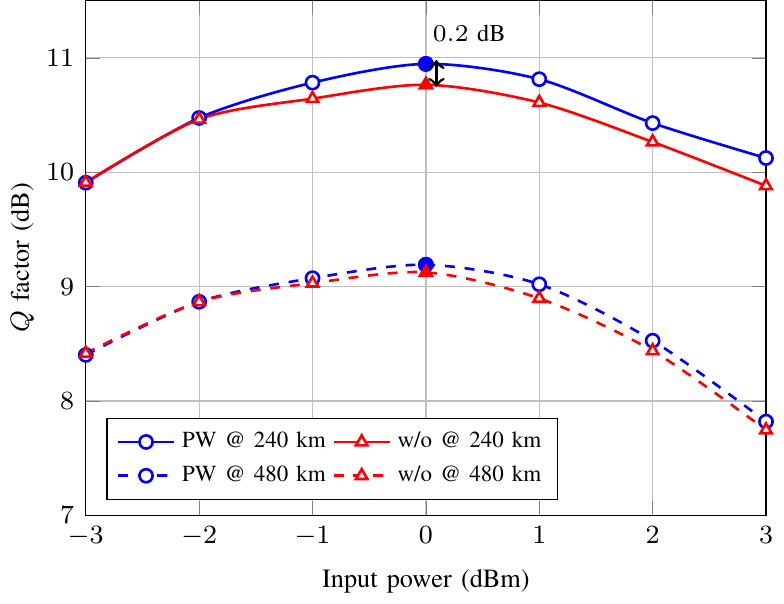}
		\caption{$Q$ factor vs. input power for DUM $64$-QAM.}	
		\label{fig:CC}
\end{figure}

We now turn our attention to the DUM system.
In Fig.~\ref{fig:BB} and Fig.~\ref{fig:CC}, a Q factor increase of about $0.4$ dB and $0.2$ dB is observed for $16$-QAM at $800$ km and $64$-QAM at $240$ km, respectively, in comparison with MD. 
At high transmission distance, PW exhibits limited improvement in comparison to the MD-based detector. This is because PW efficiently mitigates the nonlinear non-Gaussian noise by designing improved decision boundaries more adapted to the nonlinear fiber channel.  However, for uncompensated DUM system and at high transmission distance, the fiber nonlinearities behave like Gaussian noise, and are effectively modeled by the so-called Gaussian noise and enhanced Gaussian noise models \cite{gn,egn}. In this case, PW, and in general machine learning-based detectors, show limited performance improvement in comparison with the classic MD-based detector, which is the optimal detection technique for a channel with Gaussian noise.

\begin{figure}[!tbph]
		\includegraphics[width=0.93\linewidth]{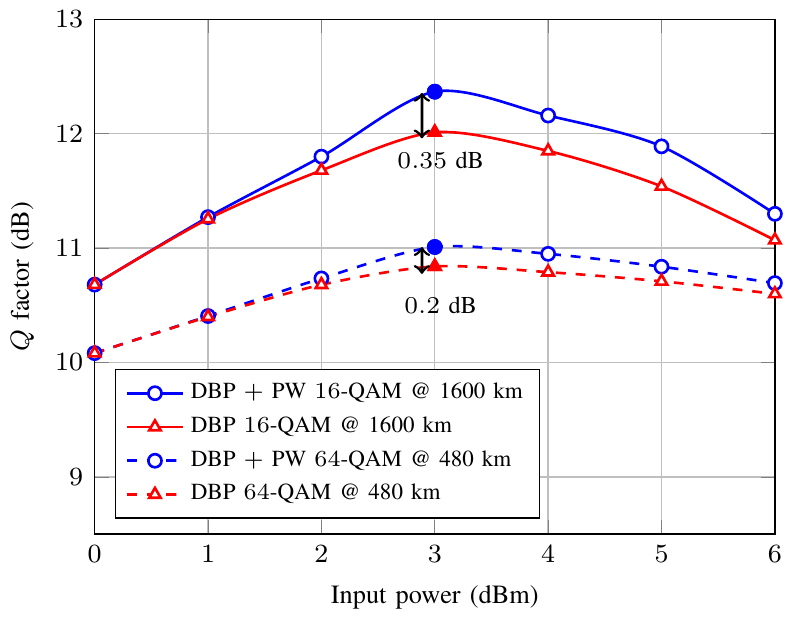}
		\caption{$Q$ factor vs. input power for DUM with DBP.}	
		\label{fig:D}
\end{figure}

 In the following, we propose a two-stage fiber nonlinearity mitigation.  Firstly, DBP is applied to compensate for the deterministic nonlinear effects. Then, PW-based detection is performed to deal with the stochastic nonlinearity due to signal-ASE noise interactions. As shown in Fig.~\ref{fig:D}, for $16$-QAM at $1600$ km and $64$-QAM at $480$ km, the two-stage compensation scheme using DBP and PW increases the performance with about $0.35$ dB  and $0.2$ dB, respectively, when compared to DBP with MD detection. This confirms that the proposed PW technique also mitigates the nondeterministic nonlinear effects due to signal-noise interactions.
 
\section{Conclusion}\label{sec:con}

We have proposed to use the Parzen window (PW) classifier as a detection technique to deal with the nonlinear non-Gaussian noise in both DM and DUM systems for different QAM modulations. Performance improvement in terms of the $Q$ factor is observed in DM systems and short reach DUM systems. This increase in performance is obtained without the need of phase rotation compensation because the detection relies on only the training data. We have also introduced a two-stage compensation using DBP and PW, which shows that PW can mitigate the stochastic nonlinear signal-ASE noise interactions, as well. An experimental validation of PW-based detector is left for future work.
 

\end{document}